\documentclass
[notitlepage,english,aps,floats,onecolumn,showpacs,nofootinbib,floatfix]
{revtex4-1}
\usepackage{pslatex}
\usepackage[T1]{fontenc}
\usepackage[latin1]{inputenc}
\usepackage{graphicx}
\usepackage{epsfig}
\usepackage{longtable}
\usepackage{float}
\usepackage{calc}
\usepackage{ifthen}
\usepackage{amsmath}

{
{
{

\newcommand{\nc}{\newcommand}
\nc{\renc}{\renewcommand}
\nc{\eqs}[2]{\mbox{Eqs.~(\ref{#1},\,\ref{#2})}}
\nc{\eq}[1]{\mbox{Eq.~(\ref{#1})}}
\nc{\figs}[2]{\mbox{Figs.~(\ref{#1},\,\ref{#2})}}
\nc{\fig}[1]{\mbox{Fig~.(\ref{#1})}}
\nc{\be}[1]{\begin{equation} \mbox{$\label{#1}$}}
\nc{\ee}{\vspace{0.1cm}\end{equation}}

\nc{\bea}[1]{\begin{eqnarray} \mbox{$\label{#1}$}}
\nc{\eea}{\vspace{0.1cm}\end{eqnarray}}

\newcommand{\bean}{\begin{eqnarray*}}
\newcommand{\eean}{\end{eqnarray*}}

\def\tdN{\tilde{N}}

\def\mso{m_{s_{0}}}

\usepackage[normalem]{ulem}

\begin{document}
\title{The 750 GeV Resonance as Non-Minimally Coupled Inflaton: Unitarity Violation and Why the Resonance is a Real Singlet Scalar}
\author{John McDonald}
\email{j.mcdonald@lancaster.ac.uk}
\affiliation{Dept. of Physics, University of 
Lancaster, Lancaster LA1 4YB, UK}

\begin{abstract}

   The 750 GeV resonance observed by ATLAS and CMS may be explained by a gauge singlet scalar. This would provide an ideal candidate for a gauge singlet scalar alternative to Higgs Inflation, known as S-inflation. Here we discuss the relevant results of S-inflation in the context of the 750 GeV resonance. In particular, we show that a singlet scalar, if it is real,  has a major advantage over the Higgs boson with regard  to unitarity violation during inflation. This is because it is possible to restrict the large non-minimal coupling required for inflation, $\xi \sim 10^5$, to the real singlet scalar, with all other scalars having $\xi \sim 1$. In this case the scale of unitarity violation $\Lambda$ is much larger than the inflaton field during inflation. This protects the inflaton effective potential from modification by the new physics or strong coupling which is necessary to restore unitarity, which would otherwise invalidate the perturbative effective potential based on Standard Model physics. This is in contrast to the case of Higgs Inflation or models based on complex singlet scalars, where the unitarity violation scale during inflation is less than or of the order of the inflaton field. Therefore if the 750 GeV resonance is the inflaton, it must be a non-minimally coupled real singlet scalar.

\end{abstract}
 
\maketitle

\section{Introduction} 

    The observation of a 750 GeV resonance at the LHC by  ATLAS \cite{atlas} and CMS \cite{cms} can be interpreted as a spin-0 gauge singlet particle. This is perhaps the most likely interpretation, although a spin-2 particle is also possible. (The interpretation of the diphoton resonance in reviewed in \cite{strumia}.) Should this be confirmed then it will the first observation of a fundamental gauge singlet scalar particle. 

  From the point of view of non-minimally coupled inflation models of the type first proposed by Salopek, Bardeen and Bond \cite{sbb}\footnote{Non-minimally coupled scalar models of inflation were also proposed in \cite{spok} and \cite{igi}. However, the scalars in these models, which correspond to induced gravity models, have large masses and expecation values in the present vacuum.}, the appearance of a second scalar in an extended Standard Model has an important implication: it provides an alternative candidate to the Higgs boson for the inflaton.  

  Gauge singlet scalars with masses in the 100 GeV-few TeV range as the basis of an alternative to Higgs Inflation \cite{hi} have been extensively studied in \cite{sinf,rlcomp,rluv,fk}, where the model was called S-inflation \cite{sinf}. (See also \cite{shafi}.) The original motivation was provided by the gauge singlet scalar dark matter model \cite{gsdm1,gsdm2,gsdm3}. The 750 GeV resonance, if it is confirmed to be a singlet scalar, will provide an alternative foundation for the model. In the original S-inflation analysis, thermal relic dark matter constraints were considered in addition to general vacuum stability and perturbativity constraints \cite{sinf,rlcomp,rluv,fk}.  However, the dark matter constraints are independent of the general S-inflation constraints and simply impose an additional restriction on the parameter space.

    A key advantage of the S-inflation model follows from unitarity violation during inflation \cite{fk,rluv}. Perturbative unitarity is violated in scalar scattering processes mediated by graviton exchange via the non-minimal coupling \cite{hertz,hmb}. If we require that the inflation model is not modified by the new physics or strong coupling required to restore unitarity then, as we will show, the inflaton scalar must be a real singlet scalar, with all other scalars having much smaller non-minimal couplings. Higgs Inflation is ruled out by this requirement, as the unitarity violation scale is of the order of the Higgs field during inflation and therefore there is no reason to expect the effective inflaton potential based on the Standard Model and perturbation theory to be valid.

The need for a TeV-scale inflaton can be considered from another perspective. It has been proposed that the naturalness of the weak scale can be understood if there are no new physics scales (in the sense of heavy particles) between the weak scale and the Planck scale \cite{shapc,lindner}. In this case, quadratic divergent corrections to the Higgs mass can be eliminated by a suitable choice of renormalization scheme and are therefore not physical. However, this would mean that all of physics needs to be explained within a TeV scale particle theory, including inflation. This strongly favours an inflation model based on a non-minimally coupled scalar field. Thus naturalness and unitarity conservation predicts  the existence of a TeV scale real gauge singlet scalar particle. This is consistent with the observation of the 750 GeV resonance. 

  The idea that the 750 GeV resonance could be due to a non-minimally coupled singlet inflaton has been discussed in \cite{raidal}, \cite{shiu} and \cite{ind}. However, these studies do not discuss unitarity violation during inflation or its implications, nor do they present a complete renormalization group (RG) analysis including the running of the non-minimal couplings and the effect of the large inflaton non-minimal coupling on the RG equations \cite{sinf,rlcomp}, which requires the inclusion of a propagator suppression factor for the inflaton \cite{clark,des}. 

    In light of the 750 GeV resonance and the need to consider the implications of unitarity violation during inflation and to correctly take into account the effect of non-minimal couplings on the RG analysis, we believe it is important to discuss the relevant results of S-inflation in this new context. In this letter we will focus on the issue of unitarity violation during inflation and explain why unitarity conservation strongly favours a real singlet scalar inflaton.

\section{Unitarity conservation strongly favours a real singlet scalar over the Higgs boson or complex singlet as the non-minimally coupled inflaton}

  The S-inflation model was originally proposed in \cite{sinf}. The most recent and complete analysis of the model is in \cite{fk}, following on from earlier studies in  \cite{rlcomp} and \cite{rluv}. As with all non-minimally coupled scalar inflation models of the form proposed in \cite{sbb}, the classical results for S-inflation are identical to those of Higgs Inflation, and are in excellent agreement with Planck,  
\be{e9}
n_s^\text{tree} \approx 1 -\frac{2}{\tdN} - \frac{3}{\tdN^2} + {\cal O}\left(\frac{1}{\tdN^3}\right) = 0.965 \;,\ee
\be{e10} r^\text{tree} \approx \frac{12}{\tdN^2}  + {\cal O}\left(\frac{1}{\xi_s \tdN^2}\right) = 3.6\times 10^{-3}  \;,\ee
Here $\tdN$ is the number of e-foldings as defined in the Einstein frame, which differs from that in the Jordan frame by $\tilde{N} \approx N + \ln(1/\sqrt{N})$~\cite{rlcomp}, and we have used $\tdN = 58$. The corresponding Planck results are $n_{s} = 0.9677 \pm 0.0060$ (68$\%$ CL, Planck TT + lowP + lensing) and $r_{0.002} < 0.11$ (95$\%$ CL, Planck TT + lowP + lensing) \cite{planckinf}. It is important to note that the Planck results point to a classical non-minimally coupled inflaton potential with at most small corrections to the potential.
It is also important to emphasize that reheating is very well defined in both S-inflation and Higgs Inflation. It occurs via preheating to SM gauge bosons in Higgs Inflation \cite{hbb} and via preheating to Higgs bosons in the case of S-inflation \cite{rlcomp}. The value of $\tilde{N}$ at the pivot scale therefore has a very small error, with $\Delta \tilde{N} \approx \pm 1$ in the case of S-inflation (corresponding to $\Delta n_s = \pm 0.001$), allowing a quite precise estimate of the inflation observables \cite{rlcomp}.

The S-inflation model is described by \cite{sinf,rlcomp,fk}
\be{e1}
 S_\text{J}  =  \int \sqrt{-g} \, \mathrm{d}^4 x \Big[ 
- \frac{m_\text{P}^2 \, R}{2} - \xi_h \, H^{\dagger}H \, R - \frac{1}{2} \, \xi_s \, s^2 \, R  + \left(\partial _\mu H\right)^{\dagger}\left(\partial^\mu H\right) + \frac{1}{2} \partial _\mu s \, \partial^\mu s - V(s^2,H^{\dagger}H) + {\cal L}_{\overline{\text{SM}}}\Big]
\;,\ee
where ${\cal L}_{\overline{\text{SM}}}$ is the SM Lagrangian density minus the purely Higgs doublet terms, $m_\text{P}$ is the reduced Planck mass and
\be{e2}
V(s^2,H^{\dagger}H) =  \left[\left(H^{\dagger}H\right) - \frac{v^2}{2}\right]^2 + \frac{\lambda_{hs}}{2} \,  s^2 \, H^{\dagger}H + \frac{\lambda_{s}}{4} \, s^4 + \frac{1}{2} \, \mso^2 \, s^2 \ee
where $v = 246 \: \text{GeV}$ is the vacuum expectation value of the Higgs field. The inflaton field during inflation is given by 
\be{e11} s_{\tilde{N}}^2 \approx \frac{4 \, m_\text{P}^2 \, \tilde{N}}{ 3 \, \xi_s} \;,\ee  while the potential along the $s$ direction in the Einstein frame during inflation is 
\be{pot}  V_{E}(\chi_s,0) = \frac{\lambda_{s} \, m_\text{P}^{4}}{4 \, \xi_s^2} \left( 1 + \exp\left(-\frac{2 \, \chi_s}{\sqrt{6} \, m_\text{P}}
\right)\right)^{-2} \;,\ee
where $\chi_{s}$ is the canonically normalized scalar field in the Einstein frame during inflation. 
The observed magnitude of the density perturbation requires that $\xi_{s}$ equals  $5 \times 10^{4} \sqrt{\lambda_{s}}$.

 In this version of S-inflation a dark matter $Z_2$ symmetry is assumed. If we generalize to the case of an unstable singlet then additional dimensionful terms are allowed, of the form $s$, $s^3$ and $s H^{\dagger}H$. However, if we assume that the mass scale of these interactions is O(TeV) then we can neglect all dimensionful terms during inflation, since inflation occurs at a field value and so renormalization scale much larger than a TeV. 
Therefore only the dimensionless quartic interactions play a role during inflation, in which case the model without a $Z_2$ symmetry is equivalent to the model with a $Z_2$ symmetry.

  We next discuss the key issue of perturbative unitarity violation. In Higgs Inflation, graviton exchange between the non-minimally coupled Higgs doublet scalars in the electroweak vacuum results in tree-level unitarity violation in high energy scalar scattering processes at\footnote{The importance of this scale was first recognized  in \cite{barbon} and \cite{burg1}.} $E  \sim  m_{p}/\xi_{h} $ \cite{hertz,hmb}. (In this it is assumed the $E$ is large enough that the electroweak gauge bosons can be considered to be massless in the scattering process.) In the inflaton background with $\overline{h} \approx \sqrt{\tdN/\xi_{h}} m_P$, the energy at which unitarity is violated in scalar scattering becomes $E \equiv \Lambda \sim  \overline{h}/\sqrt{\xi_{h}} $.  Thus perturbative unitarity breaks down at this energy. This means that either the structure of the theory changes to a unitarity conserving theory at or below this energy ('new physics'), or the problem is the breakdown of perturbation theory itself and strong coupling will unitarize the scattering rate, without requiring any modification of the theory\footnote{The latter possibility is supported by resummation of graviton propagator loops, which shows that the resummed amplitude is unitary even though the tree-level process violates unitary \cite{han,dono}.}.  

This energy scale is also less than the inflaton field and so is less than the RG-improved effective potential renormalization scale during inflation\footnote{As we will discuss, the inclusion of electroweak gauge bosons modifies the unitarity violation scale, but the conclusion remains the same.}. Therefore the calculation of the effective potential is expected to be strongly modified by the physics of unitarity conservation when the renormalization scale is comparable to the unitarity-violation scale, either by the existence of new particles\footnote{In \cite{gl} it was claimed that simply adding a singlet scalar could unitarlze Higgs Inflation. However, as explicitly demonstrated in \cite{rluv}, the resulting model is not related to Higgs Inflation, but is in fact an induced gravity inflation model in which the inflaton is a gauge singlet scalar with a mass much larger than a TeV.}  with masses less than the renormalization scale (or a more radical modification of the theory \cite{rluci}), or by perturbation theory breakdown in the computation of the quantum corrections. This means that at the renormalization scale $\mu \sim \Lambda$ there is no reason to expect the perturbative quantum effective potential based on the Standard Model to be valid. This will break the connection between low energy Standard Model physics and inflation observables and may prevent inflation if the modification of the theory and effective potential is sufficiently strong. There is therefore no reason to expect the model to be consistent with the results of Planck, since these are consistent with a classical non-minimally coupled inflaton potential with at most small corrections\footnote{The condition for the consistentcy of the perturbative effective potential is quite distinct from the condition for the calculation of scalar field fluctuations during inflation to be consistent. The latter requires that $H \ll \Lambda$, where $H$ is the typical energy associated with scalar field fluctuations. This is easily satisfied.}.

It is here that a real gauge singlet scalar has a major advantage. Unitarity violation only occurs when there are two or more non-minimally coupled scalars \cite{hertz,hmb}. This is apparent in the Einstein frame, since the single scalar model can be expressed as a conventional scalar field theory which is minimally coupled to gravity \cite{rlu1}. In the Jordan frame the absence of unitarity-violation is due to a cancellation between the s-, t- and u-channel graviton exchange scattering processes \cite{hertz}. However, in the case of two different initial state scalars this cancellation is no longer possible, as there can only be t- and u-channel diagrams in this case. 
    Following \cite{rluv} and \cite{fk}, we can consider the unitarity violation scale as a function of the inflaton field for a model with two real scalars, $\phi_{1}$ and $\phi_{2}$, with non-minimal couplings $\xi_{1}$ and $\xi_{2}$. $\phi_{1}$ is defined to be the inflaton. The action in the Einstein frame is 
\be{u1}
S_\text{E} = \int \mathrm{d}^4x\sqrt{-\tilde{g}} \left[- \frac{ m_\text{P}^2 \tilde{R}}{2}  + {\cal L}_{11} + {\cal L}_{12} + {\cal L}_{22} \,  \right],
\ee
where
\be{u2}
 {\cal L}_{ii} =  \left(\frac{\Omega^2+\frac{6 \, \xi_i^2 \, \phi_i^2}{m_\text{P}^2}}{\Omega^4}\right) \tilde{g}^{\mu\nu} \, \partial_\mu \phi_i \, \partial_\nu \phi_i ~,\ee

\be{u2a} {\cal L}_{12} = \frac{6 \, \xi_1 \, \xi_2\,\phi_1\,\phi_2 \, \tilde{g}^{\mu\nu}\,\partial_\mu \phi_1 \, \partial_\nu \phi_2}{m_\text{P}^2 \, \Omega^4}
~\ee
and 
\be{u3a} \Omega^2 = 1 + \frac{\xi_{1} \, \phi_{1}^{2} 
 + \xi_{2} \, \phi_{2}^{2} }{m_\text{P}^{2}}   \;.\ee
In this we set $V = 0$, since the potential plays no role in unitarity violation due to graviton exchange via the non-minimal coupling.  
The interaction term \eq{u2a} is 
responsible for the unitarity-violation in scattering cross-sections calculated in the Einstein frame. This interaction is the Einstein frame analogue of scalar scattering via graviton exchange in the Jordan frame due to the non-minimal coupling.

    During inflation, the inflaton has a value $\overline{\phi}_{1} = (4 \tilde{N}/3 \xi_{1})^{1/2} m_P$. 
    In this background, the interaction leading to unitarity violation in $\delta \phi_{1} \; \phi_{2}$ scattering is  
\be{vx1}  \frac{6 \, \xi_{1} \, \xi_{2}}{m_\text{P}^{2} \, \Omega^{4}} (\overline{\phi}_{1} + \delta \phi_{1}) \, \phi_{2} \,\tilde{g}^{\mu\nu} \, \partial_{\mu} \delta \phi_{1} \, \partial_{\nu} \phi_{2}   \;,\ee
where $\delta \phi_{1}$ is the fluctuation about the background inflaton field. This results in a 3-point and a 4-point interaction. The 3-point interaction produces the dominant unitarity violation \cite{rluv}. The canonically normalized fields in the Einstein frame during inflation are $\varphi_{1} = 
\sqrt{6} \, m_\text{P} \, \delta \phi_{1}/\overline{\phi}_{1}$
 and $\varphi_{2} = m_\text{P} \, \phi_{2}/(\sqrt{\xi_1} \, \overline{\phi}_{1})$. 
After rescaling to canonically normalized fields, the 3-point interaction is 
\be{vx2} \frac{\sqrt{6} \, \xi_{2}}{m_\text{P}} \varphi_{2} \, \tilde{g}^{\mu\nu} \, \partial_{\mu} \varphi_{1} \, \partial_{\nu} \varphi_{2} \;.\ee
This interaction can mediate $\varphi_{1} \varphi_{2} \leftrightarrow \varphi_{1} \varphi_{2}$ elastic scattering at energy $\tilde{E}$. The matrix element from $\varphi_2$ exchange is
\be{M_D}
{\cal M} = -\frac{6i\xi_2^2 \tilde{E}^2}{m_p^2}
~.\ee
The optical theorem condition for unitarity conservation in elastic scattering is \cite{zuber}
\be{uc1}
\left|{\rm Re}(a_l)\right| \leq \frac{1}{2}
~\ee
for all $l$, where the partial wave amplitudes $a_l$ are given by
\be{uc2}
-i{\cal M} = 16\pi\sum_{l=0}^{\infty}(2l+1)P_l(\cos\theta) a_l
~.\ee
The value of $a_0$ at tree-level is obtained by comparing \eq{uc2} to \eq{M_D}, 
\be{a0}
a_{0}^{tree} = \frac{3\xi_2^2\tilde{E}^2}{8\pi m_p^2} ~.
\ee
Applying \eq{uc1} to \eq{a0} then gives the condition for perturbative unitarity conservation in the Einstein frame 
\be{uc3}
\tilde{E} \leq \tilde{\Lambda} = \sqrt{\frac{4\pi}{3}}\frac{m_p}{\xi_2}
~. \ee
Energy scales $\tilde{E}$ in the Einstein frame are related to those in the Jordan frame by $\tilde{E} = E/\Omega$, where during inflation $\Omega^2 \simeq \xi_1 \overline{\phi}_1^2/m_\text{P}^2$. 
Therefore the perturbative unitarity violation scale in the Jordan frame is 
\be{uc3J}
\Lambda = \Omega \tilde{\Lambda} = \sqrt{\frac{4\pi}{3}} \times \frac{\sqrt{\xi_1}}{\xi_2} \overline{\phi}_{1} \sim \frac{\sqrt{\xi_{1}}}{\xi_{2}} \times \overline{\phi}_{1} 
~. \ee

The key feature of this is that if $\xi_{2} \ll \xi_{1}$ then the unitarity violation scale can be much larger than when $\xi_{2} = \xi_{1}$. The latter corresponds to Higgs inflation,  since the four real scalars in the Higgs doublet all have the same non-minimal coupling due to gauge invariance. It also correspond to a complex scalar $\Phi = \phi_{1} + i \phi_{2}$, since $\phi_{1}$ and $\phi_{2}$ have the same non-minimal coupling due to the global U(1) invariance of the complex field. When $\xi_{1} = \xi_{2}$, the unitarity violation scale is given by
\be{ua1} \Lambda \sim \frac{\overline{\phi}_{1}}{\sqrt{\xi_1}} ~,\ee 
which, with $\overline{\phi}_{1} = \overline{h}$ and $\xi_1 = \xi_h$, gives the standard result for Higgs Inflation in the inflaton background. However, since this energy scale is less than $\overline{h}$, the gauge bosons in the inflaton background are massive and decouple below this scale \cite{sb2,rluv} and only the physical Higgs scalar takes part in scattering. Since unitarity violation requires that there is more than one massless non-minimally coupled scalar, there is effectively no unitarity violation at energies less than $\overline{h}$. Unitarity violation therefore occurs at $\Lambda \approx m_{W}(\overline{h}) \approx \overline{h}$ i.e. the unitarity violation scale in Higgs Inflation is essentially equal to the Higgs field during inflation \cite{sb2}. As a result, either the new physics associated with unitarizing the theory or strong coupling effects are expected to dominate the quantum corrected effective potential during inflation.

In contrast, in the case of a real singlet scalar plus the Higgs boson we have $\phi_{1} \equiv s$ and $\phi_{2} \equiv h$. Therefore during inflation the unitarity violation scale is  
\be{ua2} \Lambda \sim \frac{\sqrt{\xi_{s}}}{\xi_{h}} \times \overline{s} ~.\ee
Therefore with $\xi_{s} \sim 10^5$ and $\xi_{h} \sim 1$, the unitarity violation scale can be a factor of 300 larger than the inflaton field during inflation\footnote{When $\xi_{s} \gg \xi_{h}$, the Einstein frame potential along the $s$ direction will be much deeper than that along the $h$ direction, since $V_{E} \propto 1/\xi^2$. Therefore it is natural in this case for inflation to occur in the $s$ direction.}.

Thus only S-inflation with a non-minimally coupled real singlet scalar can have a unitarity violation scale during inflation which is large compared to the inflaton field.  This is essential to have a consistent perturbative effective potential, which is essential in order to be sure that inflation is possible, as well as for the predictions of the model to be valid and consistent with the spectral index observed by Planck.

It should be emphasized that the results upon which this conclusion is based are all well-established in the literature. The problem of the computation of the Higgs effective potential due unitarity violation was analysed in \cite{sb2}, where it was concluded that it is not possible to compute the effective potential without full knowledge of the physics of unitarity conservation. The advantage of a singlet scalar inflaton with respect to unitarity-conservation is also well-known and easily understood in terms of the s-, t- and u-channel cancellation of graviton-mediated scalar scattering processes in the limit of a single non-miminally coupled scalar \cite{hertz}.  The purpose of the discussion we have presented here is to place these known results in the context of the possibility of a 750 GeV scalar and to make clear their implications for the nature of the scalar if it is a non-minimally coupled inflaton.

\section{Conclusions} 

   If the 750 GeV resonance is a real singlet scalar, it will provide an alternative non-minimally coupled inflaton candidate to the Higgs boson. As such, it will put S-inflation on an equal footing with Higgs Inflation as a minimal model for inflation based on known particle physics. This then raises the question of which scalar is responsible for inflation. A key requirement is that the scale of perturbative unitarity violation in scalar scattering mediated by gravitons in the inflaton background is greater than the inflaton field. If this is not the case, then the perturbative theory used to calculate the quantum effective potential in the inflaton background must be strongly modified  by the physics of unitarity conservation. It is then not justified to use the perturbative effective potential based on the Standard Model to study inflation. This will break the connection to Standard Model physics. Moreover, there is no reason to expect the unitary theory to agree with the results from Planck (which are consistent with non-minimally coupled inflation with small quantum corrections), or even to support inflation.  This perturbative unitarity requirement excludes the Higgs boson or a complex scalar as the inflaton. Only a real singlet scalar with a large non-minimal coupling, combined with small non-minimal couplings for all other scalars, has an effective potential that is consistent with perturbative unitarity during inflation. It should be emphasized that the inflaton must be a singlet with respect to any gauge interaction appearing at a mass scale less than the value of inflaton field during inflation. Therefore evidence that the 750 GeV scalar has a new beyond-the-Standard-Model gauge interaction would rule it out as the inflaton. 

  S-inflation can have observable deviations of the value of the scalar spectral index from the classical prediction, as a result of quantum corrections to the effective potential. In the original S-inflation model based on the Standard Model plus a gauge singlet scalar with Higgs portal and self couplings, $n_s - n_{s}^{tree}$ is strictly positive \cite{rluv,fk} and can easily be of the order of 0.01 for large enough values of the Higgs portal coupling $\lambda_{hs}$  (see Figure 5 of \cite{fk}). 
The 750 GeV resonance scalar is expected to have interactions with additional particles (charged and coloured scalars or vector-like fermions) which mediate its interaction with gluons and photons. These will modify the RG equations compared to the original S-inflation model. Therefore a new analysis of the RG-improved effective potential for the resonance singlet will be necessary in order to obtain its predictions for inflation observables.  

\section*{Note Added} The 750 GeV resonance has not been confirmed by the latest LHC results. However, the discussion we have presented of the advantages of the real gauge singlet scalar over the Higgs boson as a TeV-scale non-minimally coupled inflaton, as well as the prediction of such a scalar in a class of natural theories, remains valid.

\section*{Acknowledgement} 
Partly supported by STFC via the Lancaster-Manchester-Sheffield Consortium for Fundamental Physics under STFC grant ST/J000418/1.

\end{document}